# Post-*Swift* Gamma-ray Burst Science and Capabilities Needed to *EXIST*

Jonathan E. Grindlay and the *EXIST* Team

*Harvard-Smithsonian Center for Astrophysics, 60 Garden St., Cambridge, MA 02138*

**Abstract.** The exhilerating results from *Swift* in its first year of operations have opened a new era of exploration of the high energy universe. The surge to higher redshifts of the Gamma-ray bursts now imaged with increased sensitivity establishes them as viable cosmic probes of the early universe. Wide-field coded aperture imaging with solid-state pixel detectors (Cd-Zn-Te) has been also established as the optimum approach for GRB discovery and location as well as to conduct sensitive full-sky hard X-ray sky surveys. I outline the current and future major science questions likely to dominate the post-*Swift* era for GRBs and several related disciplines and the mission requirements to tackle these. The *EXIST* mission, under study for NASA's Black Hole Finder Probe (BHFP) in the Beyond Einstein Program, could achieve these objectives as the Next Generation GRB Mission with 'ultimate' sensitivity and wide-field survey capability. Analysis tools for processing *Swift*/BAT slew data are under development at CfA and will both test *EXIST* scanning imaging and provide new data on GRBs and transients.

## INTRODUCTION

In less than a decade, Gamma-ray bursts (GRBs) have moved from being one of the outstanding cosmic mysteries for which the remarkable discovery of isotropy [1] with *BATSE* laid waste earlier views of their Galactic origin, to cosmological objects, as enabled by *BeppoSAX* prompt imaging [2] which in turn allowed followup optical identifications [3] and redshifts. With *Swift* [4], GRB studies have now entered a qualitatively new phase with wide-field $\gamma$-ray imaging for prompt GRB locations accurate ($\sim 2'$) and rapid ($\sim 10$sec) enough to allow $\sim 10''$ source locations from X-ray afterglows typically only $\sim 50$-$100$sec later. As described in numerous conributions at this meeting, the questions now being asked of the *Swift* GRBs have raised the bar still further on the physics and astrophysics of GRBs, with discussion not just of star formation rates vs. redshift (SFR(z)), but metallicity vs. z (Z(z)) and – with the *Swift* first localization of a short GRB[5] (SGRB[1]) – even the rate of formation (and eventual merger) of double neutron star (DNS) systems vs. z.

Thus at each milestone in GRB research, major technological advances have, not surprisingly, spurred the next step. *BATSE* provided a large area, well-controlled (for threshold, sensitivity variations, and particle background-induced events) experiment yielding a large GRB sample, coarsely located, to provide a definitive isotropy result. *BeppoSAX* gave the first $\lesssim 10'$ GRB positions on timescales $\lesssim 0.3$d which allowed X-ray and then optical afterglows to be discovered which enabled identifications. And *Swift* has combined both prompt $\lesssim 3'$ positions and rapid re-pointing for the earliest-afterglow detections to achieve its breakthroughs for SGRB localization and identification (for which the $\sim$10-100$\times$ fainter energetics makes rapid detection essential) as well as a significantly increased mean redshift[6] (to z $\sim$2.7) enabled by its greater sensitivity to long GRBs (LGRBs).

---

[1] Both SGRB and SHB (short hard bursts) have been proposed as the de facto required acronym. I suggest SGRBs to allow for the inevitable discovery of a short soft burst – it is likely we can predict time durations better than spectral signatures for NS-NS or NS-BH merger events.

Although it is early in the *Swift* mission to make predictions for what is needed next, I nevertheless attempt to do so here so that we (collectively) may have some hope for continuing this exciting quest to the next phase(s) of discovery. I begin by outlining some of the major questions (most dealt with in far more detail by other contributions to these Proceedings) and goals for GRB studies and what they could enable in opening up the high z and high energy universe. I then outline some of the results drawn from the ongoing concept study for the Energetic X-ray Imaging Survey Telescope, *EXIST*, which could be the ultimate-sensitivity and resolution Next Generation GRB Observatory as the Black Hole Finder Probe in NASA's Beyond Einstein Program.

## OPEN QUESTIONS FOR SCIENCE OF AND ENABLED BY GRBS

In the post-*Swift* era, I suspect many of the "big" questions surrounding GRBs will still be at best only partly understood. This is likely if we take the understanding of core collapse SNe as a guide: after nearly 40y of trying, a mechanism has only just been reported[7] (but not yet confirmed) to launch a "successful" SN-II explosion and blast wave. In the following sub-sections, I describe briefly several key problems, from the (relatively) local to the most cosmologically extreme. This list is very incomplete, and represents some GRB challenges/opportunities that may have broadest impact on other disciplines. In each, I mention necessary (though perhaps not sufficient) requirements for next-generation GRB studies to make significant breakthroughs.

### Short GRBs: Unravelling the mix

While the *Swift* [5] and *HETE-2* [8] detections and first localizations of SGRBs have very likely established that they are not core collapse events (like LGRBs) and that their hosts are in elliptical or cluster galaxies within limited (December, 2006) statistics of 3-4 of 5 plausible associations (see[9] for a current summary), a number of key issues remain.

*First,* is the longstanding possible mixing of distant SGR giant flare events and SGRBs. The plausible optical hosts for the *Swift* and *HETE-2* SGRBs are too distant for SGRs (even the most luminous SGR1806-20 event of December 27, 2004, is basically undetectable by *Swift* /BAT beyond $\sim$70Mpc[10]). However SGRs could well be mixed into the *BATSE* sample of SGRBs, as suggested[11] for GRB970110. We note, however, this 13.8sec periodicity could also be due to a low frequency QPO from CygX-1 (just outside the *BATSE* position) since this would be consistent with QPOs seen at the $v_b \sim 0.1$Hz frequency characteristic[12] of CygX-1 in its low/hard state which described the source in early January 1997. Detection of the SGR component to SGRBs (which must be non-zero) will require enough sensitivity to detect the "smoking gun" magnetar pulsations, post-flash, which would not have been detectable with *BATSE* given the flux ratio for the SGR1806 event but which would be out to ∼10-40Mpc with *Swift* [10]. Extragalactic magnetar surveys would be enhanced by greater sensitivity at $\lesssim 10$ keV.

*Second,* and perhaps most interesting, are the questions now raised about SGRBs as, predominantly, DNS merger events in old stellar systems. We have suggested[13] that ∼10-30% of the SGRB population could arise from DNS systems formed by dynamical exchanges in globular cluster cores of NSs into compact binaries composed of NS-main sequence or, most often, millisecond pulsars with HeWD companions as now seen in great numbers in globular clusters. Two key observations from *Swift* and *HETE-2*, as well as the long-known DNS in a Galactic globular (M15-C) which *will* produce a SGRB, drove us to this model: 1) that the hosts are ellipticals or in galaxy clusters (with a high preponderance of ellipticals), and 2) that there is evidence[9] the DNS mergers (if indeed this is the underlying SGRB source) occurred in the (near-) haloes of their parent galaxies. What seems to not be appreciated by many in the GRB community is that offset SGRBs are *not* expected from dynamical kicks imparted to DNS systems by their NS formation events. Why? Because the 7 DNS systems known in the Galactic disk all appear to have velocities $\lesssim 50$ km s$-1$, which is consistent with recent understanding of binary evolution of DNS systems from massive binaries in which the second NS forms from a He star[14]. Although SGRBs must surely be produced by DNS mergers of these disk-bound systems, the globular cluster origin of some SGRBs will be strongly favored if *Swift* (and future missions) continue to find a signficant fraction offset from their host galaxies. Given the early evidence that SGRB afterglows are faint (no absorption line redshifts are yet detected from the SGRB afterglow events; rather all have come from the (nearby) putative host galaxy[9]), the requirement is for rapid and very accurate location of the SGRBs themselves. With higher sensitivity, energy bandwidth and resolution, the globular cluster component of the SGRB population can be isolated and used to trace the formation epoch of globulars as well as their dynamical evolution.

*Third,* and equally interesting, is the admixture of NS-NS vs. NS-BH merger events. Can the delayed "flares" in some events (e.g. the longer/softer but luminous afterpulse[8] in GRB050709 in a blue dwarf galaxy) be delayed inspiral from BH disruption of a NS? Or is it simply an afterglow from a local ISM, which for other SGRBs in or near ellipticals appear to be faint. Are the NS-BH binaries, which must exist, given kicks (despite their larger total mass than DNS systems) and thus sometimes expelled from galactic disks? Isolating the GRB signatures of NS-BH events would demonstrate that such systems exist, with their unequaled opportunity for testing GR with a precise clock (a millisecond pulsar) in close orbit around a stellar mass BH. A MSP-BH system remains a holy grail, and may be found first by MSP surveys in the radio, but the SGRB population will provide important constraints. Obviously a much larger sample of events and positions relative to their host galaxies is needed, together with ∼3-30 keV prompt spectra to look for softer afterpulses.

# Highest redshift GRBs from Pop III stellar mass BHs?

The *Swift* discovery[15] of the current record redshift GRB, at z = 6.29, is a harbinger of things to come. As discussed many times, the rapid rise of the highest z value over the past few years makes it plausible that GRBs will soon be the highest z probe of stellar objects (including galaxies themselves and their central AGN) directly observed in the universe. The long-standing goal remains to detect Pop III stars, which are likely to be massive and may produce LGRBs without quenching their jets[16]. As pointed out by Bromm and Loeb, a "successful" Pop III GRB is even more likely if the progenitor is a binary that undergoes mass loss prior to collapse. Although the predicted rate of such GRBs is very uncertain, and dependent on timescales for reionization, the Pop III GRB rate predicted for *Swift* might be $\sim$0.8 y$^{-1}$. This may be measurable with *Swift*, but clearly a larger area and field of view (FoV) future mission could increase this dramatically. The prospect of directly detecting the very first stars remains perhaps the most exciting long-term GRB objective.

Regardless of the uncertainties of Pop III GRBs, the "guaranteed" detection of Pop II GRBs (i.e. GRBs from collapsars which themselves were Pop II stars) is a prime objective. The prospect of GRBs as the "backlight" for surveying the IGM (and structures) at z $\gtrsim$7 (as practically now reached) is potentially the most dramatic use of GRBs. To make this the powerful tool that it could be requires, once again, the sample size be increased dramatically from the estimate[16] that $\sim$10% of *Swift* GRBs (or $\sim$10 y$^{-1}$) will originate from z $\gtrsim$5. Given that every such GRB is crucial, and that only $\sim$40% of all *Swift* GRBs thus far have redshift measures, it is important to to confront the *z determination efficiency problem, zdep,* to exploit the potential of GRBs for cosmology in planning for future GRB science and missions. A partial answer is contained in the reasons to measure a very large sample of GRBs, as contained in our third "Open question".

# Measuring cosmic physics with GRBs

Although much has been deciphered about the basic physics of GRBs, we really only have broad outlines of understanding of these remarkable events. Given their inherent interest, as well as great utility as probes of cosmic structure, there is great incentive to understand them by vastly larger numbers of GRBs observed with broader bandwidth, higher resolution measurements. A few examples can make the point:

*Photometric redshifts:* Much would be gained if we could remove remaining systematics from the currently-best correlative relations for GRB luminosity vs. measureable quantities – e.g. the Ghirlanda[17] relation for $E_{peak}$ vs. $E_{iso}$. By calibrating this relation, or similar variants, with a large sample (several thousand) vs. the small numbers of (sometimes differing, or at least uncertain) objects presently, we might break the bonds of *zdep*. After all, our optical colleagues did this long ago with narrow-band photometry to identify "Lyman-break" galaxies. The imperative for distance measure, and "prompt" redshift estimates cannot be overstated: the rare Pop III GRB must be observed with moderate resolution $\gtrsim$2-10micron spectra (i.e., ideally rest-frame Lyman-$\alpha$ to near-IR, to counter absorption in the IGM) to measure – ideally – clustering of absorption systems revealing cosmic structure along the line of sight. All of this would best be done within the first ∼20min of the GRB, or during its luminous prompt phase, when it could be done with perhaps a dedicated space-borne ∼1m telescope. If the prompt emission is missed, absorption line redshift measures from the much fainter afterglow would require signficant exposure time with JWST at the likely $\gtrsim$0.5d delay needed to re-point JWST. Given this (very expensive) overhead, the requirement for near-certain photometric z values for every GRB becomes obvious: there must be $\gtrsim$90% confidence that a JWST re-point is justified! Far better to hope that a co-orbiting (with *EXIST* ; see below) modest mid-IR telescope of small aperture could do prompt GRB redshift observations on *every* GRB.

*Role of magnetic fields:* The fundamental questions surrounding the GRB emission processes, and whether magnetic fields are dominant or GRB jets are Poynting flux dominated, could be revealed by polarimetry measures. The controversial claim [18] for polarization in GRB021606 does not mean that polarization is ruled out; simply that it has not yet been measured with sufficient sensitivity. Prompt vs. afterglow emission models (synchrotron vs. inverse Compton) would be possible to test and new constraints on beaming derived from polarization vs. time in the burst decay. Polarization can be measured from finely-pixellated detectors recording the spatial anisotropy of Compton scattered (∼100-500 keV) photons, for a broad range of $E_{peak}$, in high time resolution spectra.

*Nature of particle acceleration:* High energy and high time resolution spectra of the fast spikes resolved in some GRBs would reveal the nature of particle acceleration in the internal shocks thought to comprise the prompt emission. This in turn would further constrain models[19] for production of ultra-high energy cosmic rays in GRBs and is possible to measure with sufficiently large area detectors to achieve good S/N within individual spikes.

*Tests of Lorentz violations with GRBs:* The same high time resolution spectra allow tests of fundamental physics – namely models which violate Lorentz invariance and would predict dispersive or an energy dependent speed of light. Such considerations, as recently discussed by Martinez and Piran[20], would again require very sensitive high time and spectral resolution measures over a broad energy band of high-z GRBs and could be achieved with the same requirements noted above.

## *EXIST* : A NEXT GENERATION GRB CONCEPT

To answer or constrain the open questions outlined above requires a next generation GRB mission with very large collecting area and sensitivity over a large field of view and energy band. These requirements are well matched to those of the *EXIST* (Energetic X-ray Imaging Survey Telescope) all-sky deep hard X-ray imaging survey for black holes, which was recommended as one of the 3 high energy missions in the 2000 Decadal Survey Report. Over the past two years, a large collaboration has undertaken a concept study for *EXIST* as the implementation of the Black Hole Finder Probe (BHFP), one of three *Einstein Probe* missions proposed for NASA's Beyond Einstein Program. This ongoing study follows earlier studies in which the *EXIST* concept was first formulated

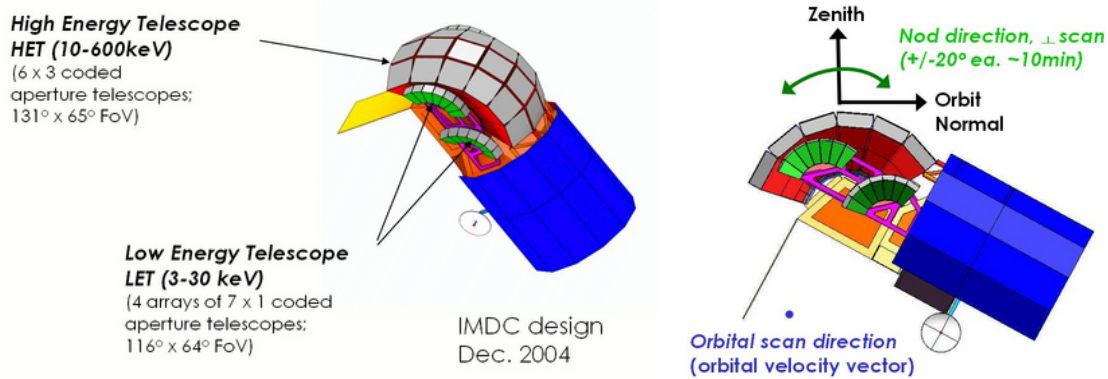

**FIGURE 1.** *Left:* Baseline concept for *EXIST* as developed at GSFC in December 2004. The large (blue) fixed solar panel array and yellow radiator panel are folded down after deployment on orbit. Scale: solar panel cylinder diameter is 5m, and the individual coded masks for the HET are 1.2m × 1.2m. *Right:* Orbital scanning configuration.

(see [21] and references therein). A detailed description of *EXIST* science generally and the full mission concept is given elsewhere[22]

The current baseline mission concept for *EXIST* is shown in Figure 1 and is centered around a very large area ($\sim$5.6m$^2$) array of imaging Cd-Zn-Te (CZT) detectors in 18 coded aperture sub-telescopes imaging a fully-coded 131° × 65° FoV. This High Energy Telescope (HET) images the full sky each orbit with 5$'$ angular resolution and $\lesssim$1$'$ source locations over the 10-600 keV band. Each HET sub-telescope includes a 56cm × 56cm array of imaging CZT (1.2mm pixels) collimated by surrounding (and rear) active shields (CsI) to a 22° × 22° fully-coded FoV defined by a radial-hole coded aperture mask. A co-aligned array of 1.1m$^2$ (total) of imaging Si detectors distributed in 28 coded-aperture sub-telescopes (20cm × 20cm each; 0.2mm pixels) constitutes the Low Energy Telescope (LET). With nearly the same FoV, the LET surveys the sky at 3-30 keV with 5× finer angular resolution and source location precision (1$'$ and $\sim$10$''$, respectively) to allow (nearly) unambiguous identifications for sources at the survey sensitivity limits. Sources a factor of (only) 3 brighter than the survey limits would be located to aspect-constrained limits of 3$''$ for unique identifications. The 5$\sigma$ sensitivity limits for *EXIST* are 2mCrab and 0.05mCrab (2 × 10$^{-11}$ and 5 × 10$^{-13}$ erg cm$^{-2}$ s$^{-1}$) for exposure times of 1 orbit (20min on source each 95min) and a 1 year survey, respectively, in any factor of 2 energy band up to 200 keV. The HET sensitivity falls off by a factor of 10 at the high energy limit (600 keV) due to both the 5mm thick CZT detectors and coded masks becoming transparent.

*EXIST* achieves full sky coverage each orbit by its zenith-pointed fan beam nodding ±20-25° each $\sim$10min (see Figure 1) so that a given source is scanned in both the orbit-velocity and orbit-normal directions for the $\sim$20min that it traverses the FoV of the telescope array. This scanning is the key to much larger dynamic range and limiting sensitivity that can extend significantly the limiting flux sensitivity below the systematic limits previously achieved in coded aperture imaging[23], as shown by simulations conducted[24] and on going as well as scanning experiments and analysis of *Swift*/BAT slew data (see below). An even more uniform sky coverage (not as overexposed at the orbit poles) is achieved by a more symmetric telescope layout, which also allows much smaller articulated solar panels, as shown in Figure 2. This new design also better matches the LET total area, and thus sensitivity, to that of the HET. The ratio is nearly 1:4, or what is required for comparable S/N per unit time for a source with a Crab-like power law spectrum (with photon index 2) in the presence of a (much) flatter cosmic diffuse background, which dominates the aperture flux below 100 keV.

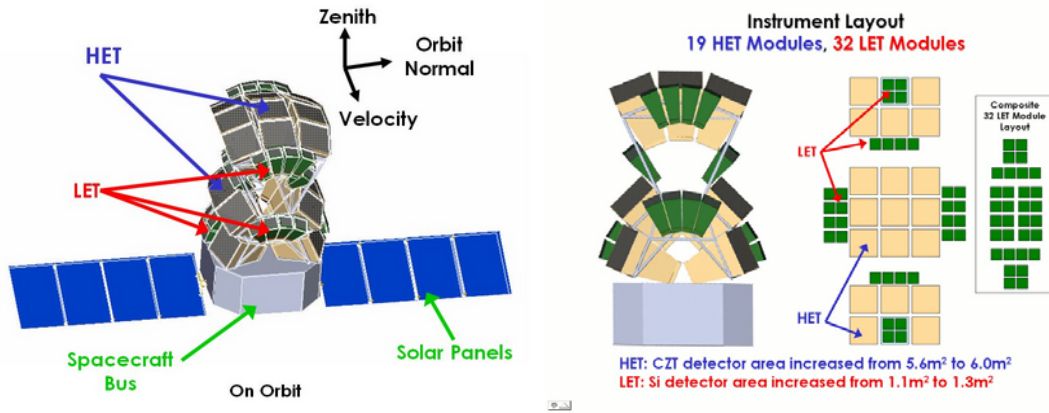

**FIGURE 2.** *Left:* New layout concept for *EXIST* as developed at General Dynamics in December 2005. The HET and LET telescope arrays are now symmetric on the S/C. *Right:* Increased fully-coded fields of view of HET (154° × 65°) and LET (160° × 64°) and total detector areas achieve more uniform sky coverage at the cost of a modest increase in total instrument mass.

## GRB sensitivities for *EXIST*

With this factor of ∼12× increase in total imaging CZT detector area over that of *Swift*/BAT, extended energy band (to both significantly lower (3keV) and higher (600 keV) energies), and larger total FoV (nearly ∼6sr, or half-sky, for $\gtrsim$50% coding response, or about 3× *Swift*), *EXIST* would achieve ∼5× greater sensitivity for GRBs and ∼10-20× greater sensitivity for its full-sky survey. Detailed simulations are being carried out over the coming year to include full background on orbit as well as effects of bright sources in the FoV (near the Galactic plane). Initial estimates by D. Band for the GRB sensitivity for *EXIST*, *Swift* and *BATSE* are shown in Figure 3 for GRBs with differing $E_{peak}$ and redshift (all referenced to the flux values shown for a GRB at z = 1). The *EXIST* sensitivity curve is plotted for a *single* sub-telescope (i.e. 1 of the 19 in the new HET configuration). A given GRB is effectively imaged over 4 sub-telescopes, so the full sensitivity reaches a factor of 2 lower flux than shown.

The active shields (1cm CsI on 4 sides of each sub-telescope around the CZT array; 2cm planar shield beneath the CZT array) are read out for GRB spectroscopy. Simulations[25] show that their sensitivity for GRB spectra nicely extends that of the CZT array to much higher energies, as shown in Figure 3. $E_{peak}$ values, and thus constraints on GRB luminosities, could be measured up to 5-10 MeV.

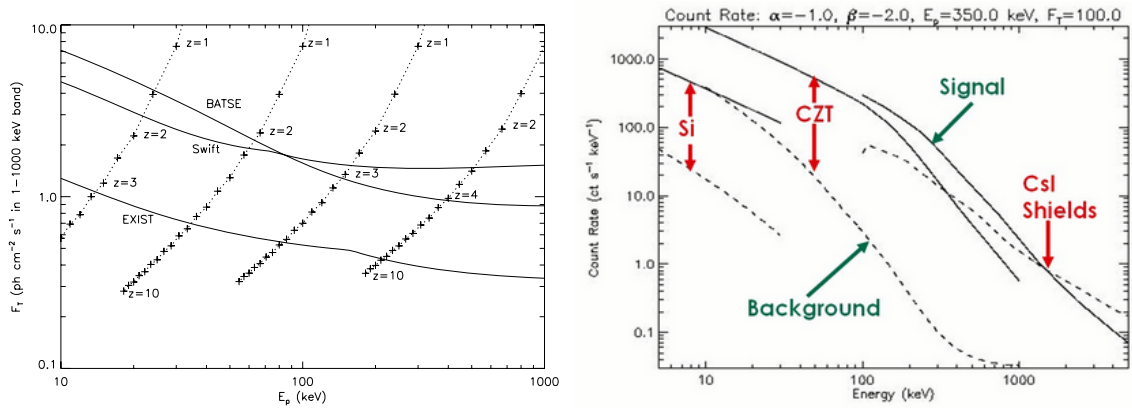

**FIGURE 3.** *Left:* Approximate sensitivity for one *EXIST* sub-telescope vs. *Swift* and *BATSE* for GRBs of $E_{peak}$ and redshift shown. Full *EXIST* GRB sensitivity would be factor of 2 better (lower flux). *Right:* Matched GRB sensitivities in the LET, HET and CsI shields for the HET for GRB with spectral parameters shown.

What does this mean for the total number of GRBs that should be detected per year by *EXIST*, or the redshift distribution of GRBs expected? Using the original analysis of Bromm and Loeb[26] gives the relative GRB numbers vs. z shown in Figure 4 for *BATSE*, *Swift* and *EXIST*. The updated estimates[16] would move the *Swift* curve closer to the *BATSE* curve; and additional corrections are indicated for the different assumptions of reionization epoch as well as Pop I/II contributions. A more complete accounting of the *EXIST* sensitivities is also warranted. Nevertheless it appears *EXIST* would detect nearly half of all cosmic GRBs at z $\sim$10 and a quarter at z $\sim$15. Thus it appears feasible to design a next generation GRB observatory to achieve close to ultimate sensitivity.

The actual rate of GRBs anticipated for *EXIST* would be $\sim$600 y$^{-1}$. Again, this estimate is being refined and may increase for more accurate treatment of full vs. partial coding in the burst imaging. GRB triggers would be derived from rate increases as well as on-board imaging. The latter would be, as for *Swift*, essential for prompt triggers on high z bursts and the sought Pop III events. This objective, as well as the general black hole sky survey objectives, require that full sky-orbit images are derived continuously each orbit. The duty cycle for recording the long-duration high z GRBs would be limited by the time over which any source is fully or partially (50%) coded: $\sim$17min or 23min, respectively, each 95min orbit, or duty cycles of 18% and 24%. Thus the scanning geometry of *EXIST* is well suited to maximizing the detection probability of very long GRBs over the full sky.

## BAT SLEW SURVEY: SCIENCE AND TESTING FOR *EXIST*

By continuous scanning its large FoV over the full sky each orbit, *EXIST* is effectively averaging over pixel to pixel systematics (e.g. gain variations, dead pixels, and changing backgrounds on any given pixel from other source variations or particle background fluctuations). Our original suggestion[23] for scanning coded apertures with radial mask holes to both minimize systematics and flatten the wide-field imaging response has been borne out by simulations[24] and laboratory tests[27]. However an actual demonstration or test of the effects of scanning can be conducted with the *Swift*/BAT coded aperture telescope by analysis of its data obtained during slews (e.g. from a fixed pointing to a GRB; or between pointed targets). We have initiated a program to develop an analysis system for BATslew data to both test coded aperture scanning and to extend the *Swift* all sky survey for persistent sources, transients, and additional GRBs. A full report on the BATslew analysis system and initial results is in preparation[28], and an early look at some first results is given here.

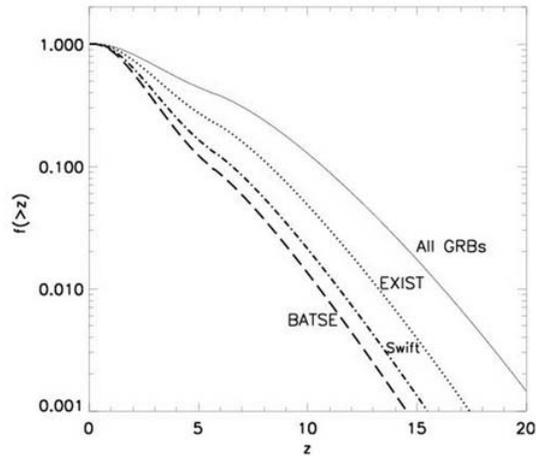

**FIGURE 4.** Estimated relative fractions of GRBs detected vs. redshift for *BATSE*, *Swift*, and *EXIST* as derived from the original formalism of Bromm and Loeb[26].

Figure 5 shows a particularly interesting BAT slew image (right) and the 76 sec pointed observation (left) that preceeded this before a final slew onto SGR050904, the $z = 6.29$ GRB[15]. The portion of the slew shown in the image (right) is truncated to be the same exposure time (76sec) as in the immediately preceeding pointing, which had (fortuitously) the bright source CygX-1 moderately near the center of the FoV of the BAT (with 46% coding fraction). The slew image has been analyzed by co-adding images each 0.2sec (the *Swift* aspect readout resolution) with proper tangent plane projections (into $10'$ sky pixels) to be centered on CygX-1. The coding fraction is larger for CygX-1 in this combined slew image, with mean value 76% vs. 47% for the pointed image. However each of the images co-added has been corrected for its (differing) coding fraction so that the total may be compared directly for source strength and S/N with the pointed image. To facilitate a BATslew survey, *Swift* is now downloading $\sim 50$ slews per day. This rich survey dataset to be explored will complement and extend the RXTE/ASM survey and recently inaugurated XMM slew survey[29].

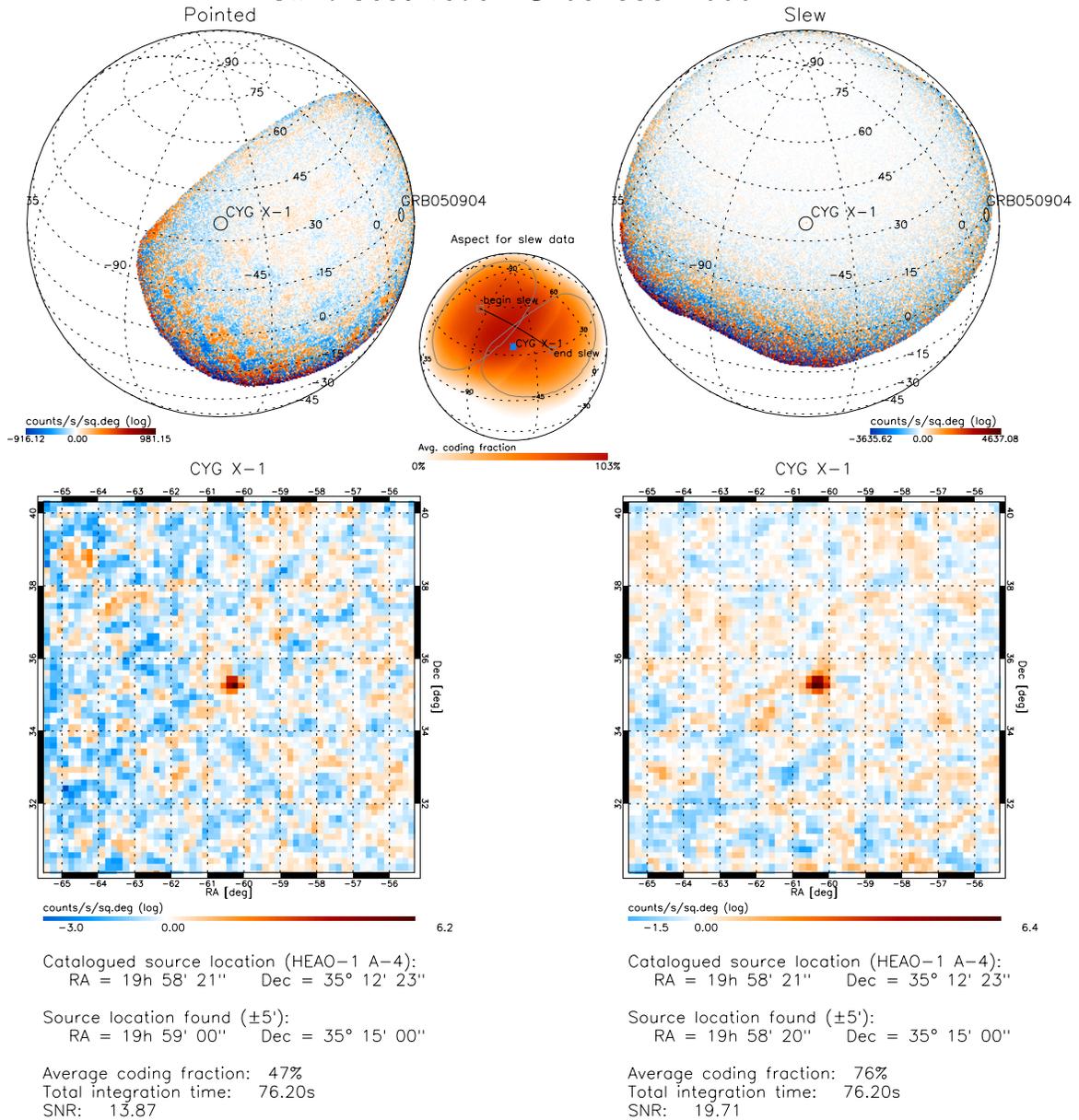

**FIGURE 5.** *Left:* Pointed image with CygX-1 near center of FoV showing full-sky view (top) and a 10° × 10° image around CygX-1 (bottom). *Right:* Same but for equal-exposure (76sec) of slew data, for which 380 × 0.2sec separate images have been co-added and weighted by their individual pixel variance values. For the slew image (right), pixels are included in the summed image only if their net exposure was $\gtrsim$10% of the maximum (i.e. $\gtrsim$7.6sec) in order to limit the effects of large fluctuations from low exposure pixels at the edge of the FoV. The resulting S/N is improved by ∼42% due to the reduced noise in the slew image vs. the pointed image (compare both the full sky as well as zoomed images; the zoomed image shows negative fluctuations only ∼half as large as the pointed observation) while the measured source flux (approximately constant between the two images) is preserved. The scan slew track is shown in the small plot centered at top: this slew was onto a target observed for only 76sec just before the subsequent slew to GRB050904, the *Swift* GRB at record z= 6.29!

# CONCLUSIONS

It is clear that GRB science is pushing the frontiers of high energy astrophysics and fundamental science, generally. As made clear at this Symposium, *Swift* is opening new vistas into the high redshift and high energy universe. It is now time to push for the Next Generation GRB mission that all of physics and astronomy should expect to make the next quantum leap. The deep questions can be answered, or better constrained, by the very large area and field of view imaging spectrometer that is required to *EXIST* . We collectively await the Beyond Einstein program to enable this next step which is, truly, Beyond Einstein.

# ACKNOWLEDGMENTS

I thank D. Band, A. Copete, D. Conte, G. Fishman, N. Gehrels, J. Hong, H. Krawczynski and G. Skinner of the *EXIST* team for contributions. This work was supported in part by NASA grant NNG04GK33G.